\documentclass[twocolumn]{aastex63} 
\usepackage{amsmath}
\usepackage[caption=false]{subfig}
\usepackage{graphicx}
\usepackage{multirow}
\usepackage{graphicx}
\usepackage{booktabs}
\usepackage{amsmath,amssymb}
\usepackage[T1]{fontenc}
\usepackage{epstopdf}
\usepackage[figuresright]{rotating}
\graphicspath{{./}{figures/}}

\begin{document}

\title{X-Ray Polarization Variability of High Spectral Peak BL Lacertaes: Cases of 1ES 1959+650 and PKS 2155--304}

\correspondingauthor{Jin Zhang \& Xiang-Gao Wang}
\email{j.zhang@bit.edu.cn; wangxg@gxu.edu.cn}

\author{Xin-Ke Hu}
\affiliation{Guangxi Key Laboratory for Relativistic Astrophysics, School of Physical Science and Technology, Guangxi University, Nanning 530004, People's Republic of China}

\author{Yu-Wei Yu}
\affiliation{School of Physics, Beijing Institute of Technology, Beijing 100081, People's Republic of China}

\author{Jin Zhang\dag}
\affiliation{School of Physics, Beijing Institute of Technology, Beijing 100081, People's Republic of China}

\author{Tan-Zheng Wu}
\affiliation{School of Physics, Beijing Institute of Technology, Beijing 100081, People's Republic of China}

\author{Ji-Shun Lian}
\affiliation{School of Physics, Beijing Institute of Technology, Beijing 100081, People's Republic of China}

\author{Xiang-Gao Wang\dag\dag}
\affiliation{Guangxi Key Laboratory for Relativistic Astrophysics, School of Physical Science and Technology, Guangxi University, Nanning 530004, People's Republic of China}

\author{Hai-Ming Zhang}
\affiliation{School of Astronomy and Space Science, Nanjing University, Nanjing 210023, People's Republic of China}

\author{En-Wei Liang}
\affiliation{Guangxi Key Laboratory for Relativistic Astrophysics, School of Physical Science and Technology, Guangxi University, Nanning 530004, People's Republic of China}

\begin{abstract}

The high-energy-peaked BL Lacertae objects (HBLs) are the main targets of the Imaging X-ray Polarimetry Explorer (IXPE) for investigating the mechanisms of radiation and particle acceleration in jets. In this paper, we report the first IXPE observations of two HBLs, 1ES 1959+650 and PKS 2155--304. Both sources exhibit X-ray polarization with a confidence level exceeding 99\%, as well as significant variability in polarization across different time intervals and energy ranges. Notably, PKS 2155--304 demonstrates the highest X-ray polarization among all blazars detected by IXPE within its entire energy band (2--8 keV), with a polarization degree of $\Pi_{\rm X}=21.9\%\pm1.9\%$ (MDP$_{99}\sim$6.0\%). An even higher polarization is observed in the 3--4 keV band, reaching $\Pi_{\rm X}=28.6\%\pm2.7\%$ (MDP$_{99}\sim$8.1\%) with a confidence level of 10.8$\sigma$. Furthermore, no polarization is detected above 5 keV energy band. For 1ES 1959+650, the highest detected polarization degree in the 2--8 keV band is $\Pi_{\rm X}=12.4\%\pm0.7\%$ (MDP$_{99}\sim$2.2\%), with an electric vector position angle (EVPA) of $\psi_{\rm X}=19.7\degr\pm1.6\degr$. The X-ray polarization of 1ES 1959+650 exhibits evident variability, accompanied by the variations of $\psi_{\rm X}$, flux, spectrum, and energy bin. We discuss possible implications of these observational findings, including the variability in polarization, rotation of EVPA, and transition between synchrotron and synchrotron-self-Compton. We speculate that the X-rays observed during different IXPE observations originate from distinct regions in the jet and may involve diverse mechanisms for particle acceleration.

\end{abstract}


\section{Introduction}

Blazars, a sub-sample of radio-loud active galactic nuclei (AGNs), are divided into BL Lacertae objects (BL Lacs) and flat spectrum radio quasars (FSRQs) based on their emission line features \citep[e.g.,][]{1995PASP..107..803U}. Currently, the detected extragalactic TeV sources mainly consist of blazars, specifically BL Lacs. Consequently, BL Lacs serve as the main targets for studying the cosmic $\gamma$-ray horizon \citep[e.g.,][]{2013ApJ...770...77D} and the intergalactic magnetic fields \citep[e.g.,][]{2012ApJ...747L..14V}. The broadband spectral energy distributions (SEDs) of blazars typically exhibit a double-peaked structure. The first peak is observed in the infrared--optical--ultraviolet band and even extends to X-rays, originating from synchrotron radiation of relativistic electrons. The second peak spans the GeV--TeV $\gamma$-ray regime and is believed to stem from inverse Compton (IC) scattering process of relativistic electrons \citep[see,][]{1992ApJ...397L...5M,1996MNRAS.280...67G,2009ApJ...704...38S,2012ApJ...752..157Z,2014ApJ...788..104Z,2015ApJ...807...51Z,2014ApJS..215....5K}. According to the properties observed in radio and X-ray surveys, BL Lacs were initially classified as radio-selected BL Lacs and X-ray selected BL Lacs \citep{1995PASP..107..803U}. However, a more physically motivated approach for categorizing BL Lacs is based on their peak frequencies ($\nu_{\rm s}$) of the lower-energy bump in the SED. This leads to the classification of low-energy peaked BL Lacs (LBLs) with $\nu_{\rm s}<10^{14}$ Hz, intermediate-energy peaked BL Lacs (IBLs) with $10^{14}<\nu_{\rm s}<10^{15}$ Hz, and high-energy peaked BL Lacs (HBLs) with $\nu_{\rm s}>10^{15}$ Hz. FSRQs commonly belong to low synchrotron peaked (LSP) sources, which are similar to LBLs.

Due to the high flux and violent variability in the X-ray band, blazars, particularly HBLs, are the main targets of the Imaging X-ray Polarimetry Explorer (IXPE). Up to now, IXPE has observed approximately ten blazars, among which three HBLs (Mrk 501, Mrk 421, PG 1553+113) show significant linear polarization in X-rays with the polarization degrees >$\sim$10\% \citep{2022Natur.611..677L,2022ApJ...938L...7D,2023ApJ...953L..28M}. However, none of the three LSP FSRQs (3C 273, 3C 279, 3C 454.3), along with one IBL S5 0716+714 demonstrate X-ray polarization at a significance level of 3$\sigma$
\citep{2023arXiv231011510M}. The first two IXPE observations of BL Lacertae provide only an upper limit for the polarization degree ($\Pi_{\rm X}$<12.6\% ) in the 2--8 keV band. Conversely during an X-ray burst event when it exhibited a similar X-ray spectrum to IBLs, IXPE detected a high X-ray polarization of $\Pi_{\rm X}$>20\% in the 2--4 keV band \citep{2023ApJ...948L..25P}. All these observation results from IXPE support a leptonic radiation model to explain the X-ray emission of blazars; the X-rays of HBLs are produced by the synchrotron radiation of relativistic electrons in a region with ordered magnetic fileds while the X-rays of LBLs and IBLs may be dominated by the IC process. This suggests that proton-synchrotron radiation likely does not contribute significantly.

1ES 1959+650 \citep[$z=0.047$;][]{1996ApJS..104..251P} and PKS 2155--304 \citep[$z=0.116$;][]{1993ApJ...411L..63F} are two TeV HBLs \citep{1999ApJ...513..161C,2003ApJ...583L...9H} and are also the targets of the IXPE observations. In this paper, we present their first X-ray polarimetry observations by IXPE, along with a comprehensive analysis of their long-term multiwavelength data obtained simultaneously from Swift--XRT, NuSTAR, and Fermi--LAT observatories. Section 2 provides details on the observations and data analysis methodology employed in this study. The observation results are described in Section 3, followed by conclusions and discussion presented in Section 4.
 
\section{Observations and Data Analysis}

\subsection{IXPE}

IXPE observed 1ES 1959+650 four times between 2022 May 3 and 2023 August 14 with the net exposure time ranging from $\sim$50 ks to $\sim$300 ks. Additionally, it performed one observation of PKS 2155--304 between 2023 October 27 and 2023 November 7, with an extremely long exposure time of over 470 ks, as listed in Table \ref{table1}. We analyze the publicly available level-2 event files that store polarization information in form of Stokes parameters ($I$, $Q$ and $U$) photon by photon to derive the polarization parameters for each individual observation. The source and background regions are identified using the \emph{SAOImageDS9} software \citep{2006ASPC..351..574J}. The source region is defined as a circle with a radius of 60$^{\prime\prime}$ centered on the brightest pixel. The background region is defined as an annulus with inner and outer radii of 120$^{\prime\prime}$ and 270$^{\prime\prime}$ respectively. The same strategy is applied to the observations of both 1ES 1959+650 and PKS 2155--304. 

According to the method presented in \citet{2015APh....68...45K}, we analyze the data using $ixpeobssim$ \citep{2022SoftX..1901194B}, which is powerful software customized for IXPE data analysis and simulations. Photons from the source and background regions are extracted with the $xpselect$ task. The polarization is calculated via PCUBE algorithm within $xpbinview$ task. We generate polarization cubes for the three detect units (DUs) to extract information, such as Stokes parameters, minimum detectable polarization at 99\% significance (MDP$_{99}$), polarization degree ($\Pi_{\rm X}$), electric vector position angle ($\psi_{\rm X}$), and their associated errors in the 2--8 keV band. The estimated polarization parameters from the three DUs for each IXPE observation are given in Table \ref{table1}. 

The result showing significant polarization is cross-checked by performing spectropolarimetric analysis using Xspec \citep{1999ascl.soft10005A} in the HEASoft (v.6.30.1) package. We create the Stokes parameter spectra of the source and background using the PHA1, PHAQ, and PHAU algorithms, which map the $I$, $Q$, and $U$ Stokes parameters of the photons into OGIP-compliant pulse height analyzer (PHA) files (three Stokes parameter spectra per three DUs). To apply $\chi^2$ statistics for spectropolarimetric fitting, we rebin the $I$, $Q$ and $U$ spectra using the FTOOL task $ftgrouppha$. A minimum of 30 counts per spectral channel is required for the $I$ spectra. A constant energy bin of 0.2 keV is applied to the $Q$ and $U$ spectra. We simultaneously fit 3 $\times$ $I$, $Q$ and $U$ spectra with an absorbed single power law model with a constant polarization of the form $TBabs \times (polconst \times po)$ within Xspec. The Galactic photoelectric absorption is considered in the $TBabs$ model, where the neutral hydrogen column density is fixed at Galactic value: i.e., $N_{\rm H}= 1.01\times 10^{21}$ cm$^{-2}$ for 1ES 1959+650 and $N_{\rm H} = 1.28 \times 10^{20}$ cm$^{-2}$ for PKS 2155--304, respectively \citep{2016A&A...594A.116H}. The polarization model $polconst$, assumeing constant polarization parameters within the operating energy range, has two free parameters: $\Pi_{\rm X}$ and $\psi_{\rm X}$. The results of spectropolarimetric fitting are also presented in Table \ref{table1}.

\subsection{Swift-XRT}

The X-ray Telescope \citep[XRT;][]{2005SSRv..120..165B} on board the Neil Gehrels Swift Observatory \citep[Swift;][]{2004ApJ...611.1005G} has monitored both 1ES 1959+650 and PKS 2155--304 for a long time. Being quasi-simultaneous with the IXPE observations, the XRT performed 15 observations of 1ES 1959+650 and 10 observations of PKS 2155--304 in the Windowed Timing (WT) and Photon Counting (PC) readout modes. Excluding one observation with poor data quality,  the data from 14 observations of 1ES 1959+650 and 10 observations of PKS 2155--304 are analyzed in this work. All the data are processed using the XRTDAS software package (v.3.7.0), which was developed by the ASI Space Science Data Center (SSDC) and released by the NASA High Energy Astrophysics Science Archive Research Center (HEASARC) in the HEASoft package. The calibration files from XRT CALDB (version 20220803) are used within the $xrtpipeline$ to calibrate and clean the event files. For WT readout mode data, events for spectral analysis are selected from a circle with a radius of 20 pixels ($\sim46^{\prime\prime}$) centered around the source position. For PC readout mode data, a count rate above $\sim$0.5 counts/s indicates a significant pile-up effect in the inner part of the point-spread function (PSF). Therefore, we extract events for spectral analysis from an annulus with inner and outer radii of 6 pixels and 20 pixels centered on the source position to remove the pile-up effect. The background is estimated in an annulus with inner and outer radii of 30 pixels ($\sim71^{\prime\prime}$) and 45 pixels ($\sim106^{\prime\prime}$) for both WT and PC readout modes. The ancillary response files (ARFs), which are applied to correct the PSF losses and CCD defects, are generated with the $xrtmkarf$ task using the cumulative exposure map. The spectra are grouped to ensure at least 20 counts per bin, then fitted in Xspec with an absorbed single power-law model, where we also only consider the Galactic absorption and fix the neutral hydrogen column density at Galactic value as done in IXPE spectropolarimetric analysis. The $\chi^{2}$ minimization technique is adopted for all spectral analyses. Note that a log-parabola model instead of a power-law model is needed to fit quasi-simultaneously observed spectra of 1ES 1959+650 during the fourth IXPE observation. The log-parabola function is
\begin{equation}
\frac{dN}{dE}=N_{0}(\frac{E}{E_0})^{-(\Gamma_{\rm X}+\beta{\log}(\frac{E}{E_0}))},
\end{equation}
where the decimal logarithm is used, $E_0$ is the scale parameter of photon energy, $\Gamma_{\rm X}$ is the photon spectral index, and $\beta$ is the curvature parameter \citep{2004A&A...413..489M}. If $\beta=0$, the log-parabola function turns into a single power-law function, being
\begin{equation}
\frac{dN}{dE} = N_{0}(\frac{E}{E_0})^{-\Gamma_{\rm X}}. 
\end{equation}

To investigate the X-ray emission state of 1ES 1959+650 and PKS 2155--304, we also obtain their long-term X-ray light curves from a long-term Swift monitoring program of $Fermi$ $\gamma$-ray sources\footnote{\url{https://www.swift.psu.edu/monitoring/}} \citep{2013ApJS..207...28S}, which has considered the background subtraction, PSF correction, and pile-up correction, as displayed in Figure \ref{LC}. The average count rate for each IXPE observation of the two sources is also presented in Figure \ref{LC}. Note that the data during the fourth IXPE observation of 1ES 1959+650 and during the IXPE observation of PKS 2155--304 are not available on this website. Therefore, we estimate the average count rate during these periods through our own data analysis, as indicated by the green line for 1ES 1959+650 and magenta line for PKS 2155--304 in Figure \ref{LC}.

\subsection{NuSTAR}

The data used in this work includes both FPMA and FPMB spectra from a quasi-simultaneous observation of 1ES 1959+650 by NuSTAR \citep{2013ApJ...770..103H} on October 31, 2022. We process the raw data using the script $nupipeline$ in NuSTAR Data Analysis Software (v.2.1.2) to obtain calibrated and cleaned event files. Source data are extracted from a circle with a radius of 50$^{\prime\prime}$ centered on the centroid of X-ray emission, while background is estimated in an annulus with inner and outer radii of 100$^{\prime\prime}$ and 200$^{\prime\prime}$, respectively. The spectra are grouped to ensure at least 20 counts per bin and the $\chi^{2}$ minimization technique is adopted for the spectral analysis. The spectra are produced via the NuSTAR data analysis package $nuproducts$, and then we perform the spectral fits with the standard NuSTAR response matrices and effective area files. The spectra are fitted by a log-parabola model with one absorption component; neutral hydrogen column density is fixed at Galactic value as done in IXPE data analysis. Details of the best-fit parameters are presented in Table \ref{table2}.

\subsection{Fermi--LAT}

1ES 1959+650 and PKS 2155--304 have been reported to be associated with the Fermi-LAT sources 4FGL J2000.0+6508 and 4FGL J2158.8--3013, respectively \citep{2023arXiv230712546B}. We analyze their Fermi-LAT observational data, covering over a period of 15 yr (MJD 54683--60263, from 2008 August 4 to 2023 November 15). The latest Pass 8 data are extracted from the Fermi Science Support Center for our analysis. We consider events in the energy range of 0.1--300 GeV and a region of interest (ROI) with a radius of 15$\degr$ centered at the source radio position, i.e., (R.A.= 299.999$\degr$, Decl.= 65.1485$\degr$) for 1ES 1959+650 and (R.A.= 329.717$\degr$, Decl.= -30.2256$\degr$) for PKS 2155--304. The \textit{Fermitools} (version 2.2.0)\footnote{\url{https://fermi.gsfc.nasa.gov/ssc/data/analysis/software/}} and the P8R3\_SOURCE\_V3 set of instrument response functions are used for the data analysis. A zenith angle cut of 90$\degr$ is set to avoid $\gamma$-ray contamination caused by the Earth's limb. The sources within ROI listed in the LAT Fourth Source Catalog \citep[4FGL-DR4;][]{2023arXiv230712546B} are included in the model. The spectral parameters of all sources lying within 8$\degr$ are left free, whereas the parameters of those sources lying beyond 8$\degr$ are fixed
to their 4FGL-DR4 values. The normalization parameters of the Galactic diffuse component (gll\_iem\_v07.fits) and the isotropic emission (iso\_P8R3\_SOURCE\_V3\_v1.txt) are kept free.

The photon spectra of both 1ES 1959+650 and PKS 2155--304 within the energy band of 0.1--300 GeV are well described by a log-parabola spectral function, 
\begin{equation}
\frac{dN}{dE}=N_{0}(\frac{E}{E_0})^{-(\Gamma_{\gamma}+\beta{\log}(\frac{E}{E_0}))},
\end{equation}
where $\Gamma_{\gamma}$ is the photon spectral index. We use the maximum test statistic (TS) to assess the significance of the $\gamma$-ray signals, with $\rm TS=2(log\mathcal{L}_{\rm src}-log\mathcal{L}_{\rm null})$, where $\mathcal{L}_{\rm src}$ and $\mathcal{L}_{\rm null}$ denote the likelihood values for background with and without a source. 

The light curves are obtained from the derived fluxes with the log-parabola spectral fits and extracted using an adaptive-binning method \citep{2012A&A...544A...6L}, which is based on a criterion of TS$\textgreater$25 for each time bin. The minimum time-bin size is set to be 15 days for the long-term light curve and 6 hr for the short-term (during IXPE observations) light curve. 

\section{Results}

\begin{figure}
    \centering
    \includegraphics[angle=0,scale=0.6]{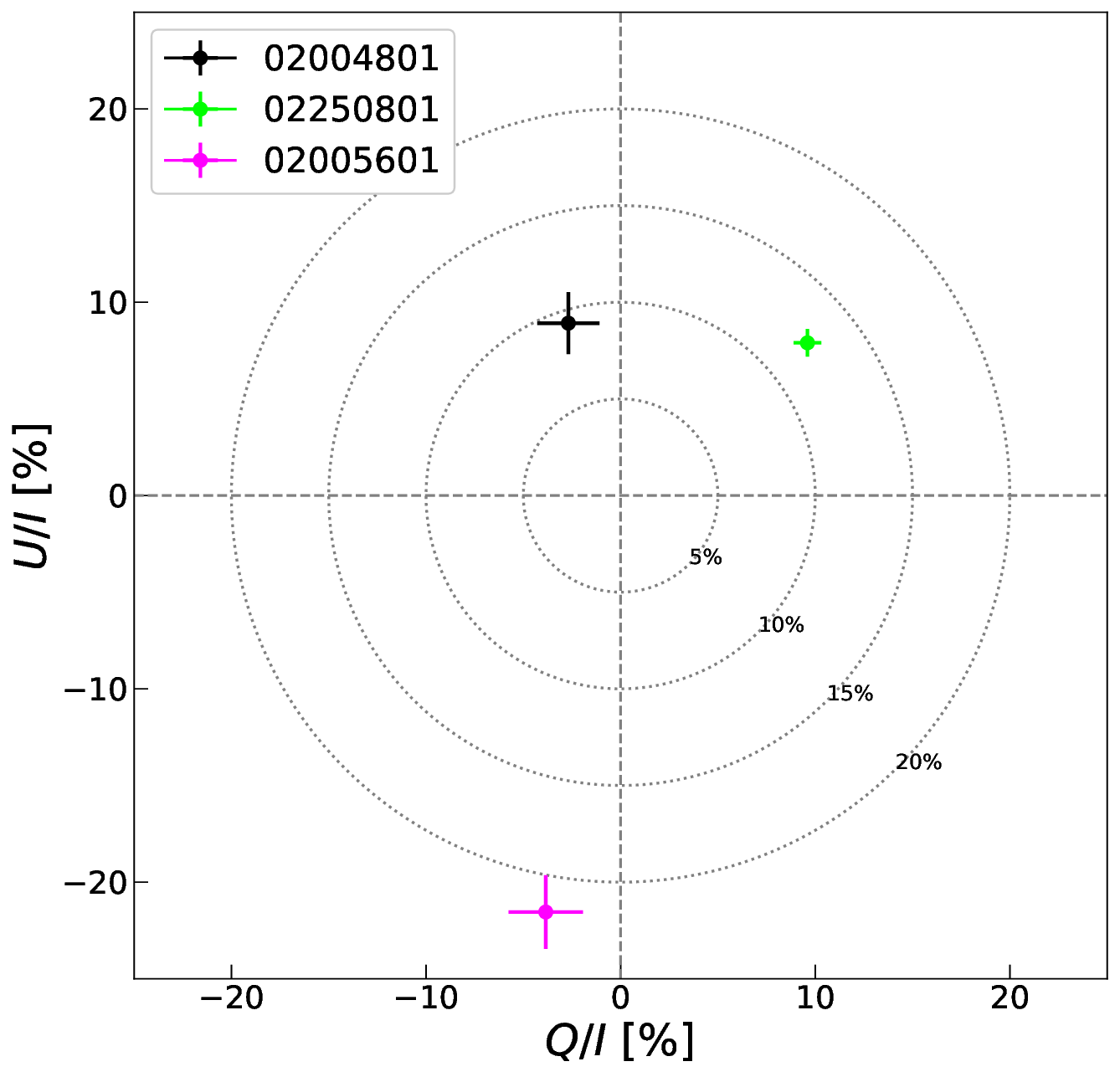}
    \caption{Normalized $Q/I$ and $U/I$ Stokes parameters in the 2--8 keV band. The black and green points represent the results of the third and fourth IXPE observations for 1ES 1959+650, respectively. The magenta point represents the result of the IXPE observation for PKS 2155--304.}\label{IXPE}
\end{figure}

We analyze the observational data from IXPE for two HBLs, namely 1ES 1959+650 and PKS 2155--304. Our analysis reveals that both sources exhibit significant polarization in the energy range of 2--8 keV, as depicted in Figure \ref{IXPE}. To investigate the energy-dependent variations in polarization, we calculate the polarization degree for each energy bin with a size of 1 keV, as shown in Figure \ref{IXPE_kev}. Two different methods are employed to estimate the polarization parameters for these two sources, yielding consistent results presented in Table \ref{table1}. Additionally, we perform an analysis on simultaneous observation data obtained from Swift--XRT, NuSTAR, and Fermi--LAT. The main findings are summarized below.

\subsection{1ES 1959+650}

Only two out of the four IXPE observations detected significant polarization in the 2--8 keV band, specifically the third and fourth observations, as indicated in Table \ref{table1}. The measured polarization parameters for these observations are $\Pi_{\rm X}=9.3\%\pm1.6\%$ with $\psi_{\rm X}=53.4\degr\pm4.9\degr$ for the third observation, and $\Pi_{\rm X}=12.4\%\pm0.7\%$ with $\psi_{\rm X}=19.7\degr \pm 1.6 \degr$ for the fourth observation, respectively. Both observations exhibit a higher level of polarization compared to their corresponding MDP$_{99}$ values (i.e., 4.9\% and 2.2\%), particularly for the fourth observation where the confidence level significantly exceeds 5$\sigma$. No significant polarization was detected in the 2--8 keV band during either the first or second IXPE observations.

However, the first IXPE observation revealed a polarization of $\Pi_{\rm X}=9.4\%\pm3.1\%$ (MDP$_{99}\sim9.3\%$) in the 2--3 keV energy range\footnote{The estimated polarization parameters are $\Pi_{\rm X}=9.4\%\pm2.3\%$ (MDP$_{99}\sim7.1\%$) and $\psi_{\rm X}=129.2\degr\pm7.1\degr$ in the 2--4 keV.}, with $\psi_{\rm X}=129.9\degr\pm9.3\degr$, as depicted in Figure \ref{IXPE_kev}. No noticeable polarization was observed across any energy bin during the second IXPE observation. Moreover, only the low-energy bands (2--3 keV, 3--4 keV, and 4--5 keV) exhibited significant polarization during the third IXPE observation. In contrast, all energy bins displayed substantial polarization during the fourth IXPE observation\footnote{We also estimated the polarization degree in energy bins of 2 keV and obtained $\Pi_{\rm X}=11.4\%\pm0.7\%$ (MDP$_{99}\sim2.1\%$), $\Pi_{\rm X}=14.0\%\pm1.6\%$ (MDP$_{99}\sim4.7\%$), and $\Pi_{\rm X}=24.0\%\pm4.6\%$ (MDP$_{99}\sim13.9\%$) for the 2--4 keV, 4--6 keV, and 6--8 keV bands, respectively.}, as shown in Figure \ref{IXPE_kev}, with a noticeable increase in polarization at higher energies; an astonishingly high level of polarization, $\Pi_{\rm X}=37.0\%\pm9.3\%$ (MDP$_{99}\sim28.0\%$), was observed in the 7--8 keV band with a confidence level of 3.6$\sigma$.

Our analysis utilizes all XRT observational data and one NuSTAR observational data obtained during the IXPE observation times, as presented in Table \ref{table2}. Figure \ref{XRT} displays one or two X-ray spectra during each IXPE observation, along with the corresponding model fitting. In the XRT data, the source exhibits obvious spectral variation: a power-law spectrum in the low-flux state and a log-parabola spectrum during bright flares. However, when combining simultaneous observational data of XRT and NuSTAR, a log-parabola model provides a better fit to the joint spectrum than a power-law model, even during a low-flux state (as illustrated in Figure \ref{XRT}). As shown in Table \ref{table2}, although the XRT flux changes by a factor of more than 4, the value of $\Gamma_{\rm X}$ remains at around $\Gamma_{\rm X}\sim2$. There is no evidence of a \emph{harder when brighter} trend observed in the 0.3--10 keV band, as seen in other blazars.

Note that the first two IXPE observations commenced on 2022 May 3 and 2022 June 9, respectively, during a period when the source is likely in an intermediate emission state in both X-ray and $\gamma$-ray bands. After four months (starting from 2022 October 28), the IXPE detected significant polarization of the source when its X-ray flux is significantly low and $\gamma$-ray flux is relatively high. Subsequently, ten months later (beginning on 2023 August 14), the IXPE once again detected significant X-ray polarization while the source is in a high X-ray flux state with intermediate $\gamma$-ray emission levels. This is illustrated in Figure \ref{LC} and can also be referred to in Tables \ref{table2} and \ref{table4}. To further investigate the correlation between X-ray and $\gamma$-ray fluxes during the IXPE observations, we derived light curves for both bands using observational data from Swift--XRT and Fermi--LAT taken simultaneously with the third and fourth IXPE observations. These are displayed in Figure \ref{LC-IXPE}. During the fourth IXPE observation period, there is evident variability observed in X-ray emission at a confidence level of 30.2$\sigma$, while $\gamma$-rays remain relatively stable. However, neither X-ray nor $\gamma$-ray fluxes exhibit significant variations during the third IXPE observation.

The X-ray polarization of 1ES 1959+650, as described above, exhibits evident variability and is accompanied by the variations of flux, spectrum, energy bin, and $\psi_{\rm X}$. Comparing with the X-ray spectra, the $\gamma$-ray spectra within the LAT band consistently require a log-parabola function for explanation; they are harder than the X-ray spectra but not as curved. According to the typical one-zone leptontic radiation model of BL Lacs, a value of $\Gamma_{\gamma}<2$ indicates that $\gamma$-rays in the 0.1--300 GeV range are produced through synchrotron self-Compton (SSC) processes \citep[e.g.,][]{1996MNRAS.280...67G,2012ApJ...752..157Z}. Therefore, it can be inferred that different electron populations contribute to the X-rays and $\gamma$-rays of 1ES 1959+650 through distinct radiation mechanisms.

\subsection{PKS 2155--304}

To date, only one observation of PKS 2155--304 has been conducted by IXPE, spanning from 2023 October 27 to 2023 November 7, with a total exposure time of 476 ks. The polarization parameters in the 2--8 keV band, estimated via the $ixpeobssim$ software, are $\Pi_{\rm X}=21.9\%\pm1.9\%$ (MDP$_{99}\sim$6.0\%) and $\psi_{\rm X}=129.9\degr\pm2.5\degr$, indicating a highly significant detection of polarization at a confidence level exceeding 5$\sigma$. Additionally, to employ spectropolarimetric fitting in Xspec yields derived values for the polarization parameters as $\Pi_{\rm X}=24.1\%\pm1.6\%$ and $\psi_{\rm X}=129.4\degr\pm1.9\degr$, as presented in Table \ref{table1}. Comparing these results with those obtained using the $ixpeobssim$ software, while the EVPA value remains nearly identical there is a slight increase in $\Pi_{\rm X}$, although both measurements are still consistent within their respective error ranges.

We estimated the polarization degree in energy bins of 1 keV and obtained $\Pi_{\rm X}$ for each energy bin, as depicted in Figure \ref{IXPE_kev}. The analysis revealed that only the low-energy energy bins (2--3 keV, 3--4 keV, and 4--5 keV) exhibit significant polarization. The estimated polarization degree in the energy bin of 3--4 keV is notably higher than those in the other two energy bins, with $\Pi_{\rm X}=28.6\%\pm2.7\%$ (MDP$_{99}\sim$8.1\%). The observed polarization is at a confidence level of 10.8$\sigma$.

As illustrated in Figure \ref{LC}, the X-ray emission of PKS 2155--304 during IXPE observation is exceptionally low, reaching its lowest state in over a decade. However, among all the blazars observed, this BL Lac exhibits the highest polarization degree in the 2--8 keV range as detected by IXPE. The XRT spectra of PKS 2155--304 (as listed in Table \ref{table3}) are generally softer than those of 1ES 1959+650, with an average photon spectral index ($\Gamma_{\rm X}$) greater than 2. This suggests that the synchrotron peak lies below the energy band covered by XRT and that IXPE captures the high-energy tail of synchrotron radiation for this source. Despite its very low X-ray flux during IXPE observation, clear flux variations are evident (Figure \ref{LC-IXPE}) at a confidence level of 4.7$\sigma$. The flux variations are accompanied by spectral changes, demonstrating a tendency towards \emph{harder when brighter} behavior within the XRT band, as shown in Figure \ref{F-G_2155}. However, no significant flux variation is observed within the LAT energy band during IXPE observation.

\section{Conclusions and Discussion}

We have reported the first IXPE observations of two HBLs, 1ES 1959+650 and PKS 2155--304, finding their significant variability in the X-ray polarization across different time intervals and energy ranges. Among the IXPE-detected blazars, PKS 2155--304 exhibits the highest X-ray polarization degree over the whole IXPE energy band (in the 2--8 keV band), with a value of $\Pi_{\rm X}=21.9\%\pm1.9\%$ (MDP$_{99}\sim$6.0\%). Notably, when considering a narrower energy bin of 1 keV, the polarization reaches an even higher level for both sources: $\Pi_{\rm X}=37.0\%\pm9.3\%$ (MDP$_{99}\sim28.0\%$) in the 7--8 keV band for 1ES 1959+650 at a confidence level of 3.6$\sigma$, and $\Pi_{\rm X}=28.6\%\pm2.7\%$ (MDP$_{99}\sim$8.1\%) in the 3--4 keV band for PKS 2155--304 at a confidence level of 10.8$\sigma$.

\begin{figure}
    \centering
    \includegraphics[angle=0,scale=0.65]{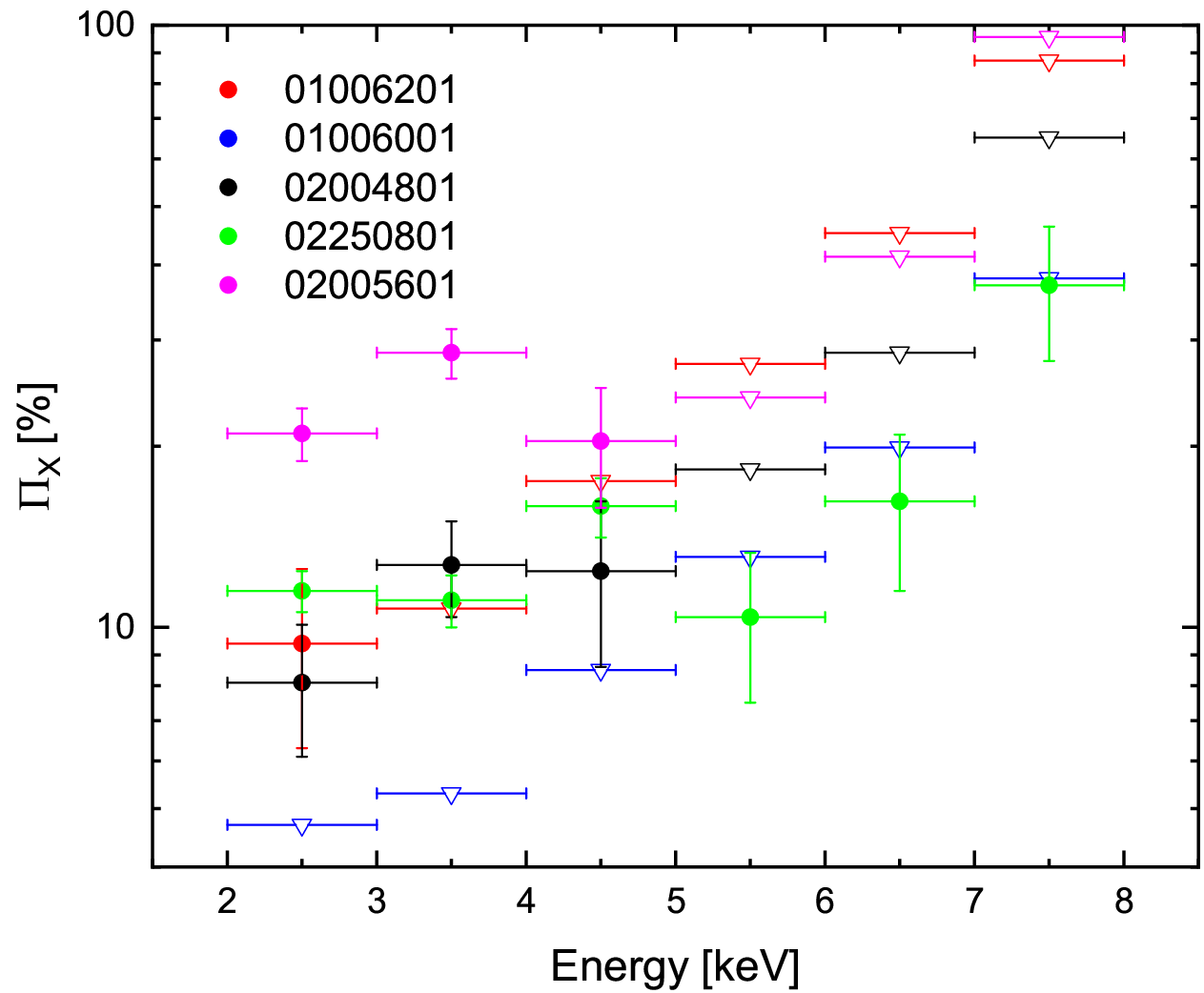}
    \caption{Polarization degree as a function of energy bin for 1ES 1959+650 and PKS 2155--304, where the polarization degrees were estimated by the software $ixpeobssim$. The red, blue, black and green points represent the results of the first, second, third, and fourth IXPE observations for 1ES 1959+650, respectively. The magenta points represent the result of the IXPE observation for PKS 2155--304. If the estimated value of $\Pi_{\rm X}$ is smaller than MDP$_{99}$, then the MDP$_{99}$ value is taken as an upper limit for that energy bin and represented as a opened inverted triangle.}\label{IXPE_kev}
\end{figure}

\begin{figure}
    \centering
    \includegraphics[angle=0,scale=0.4]{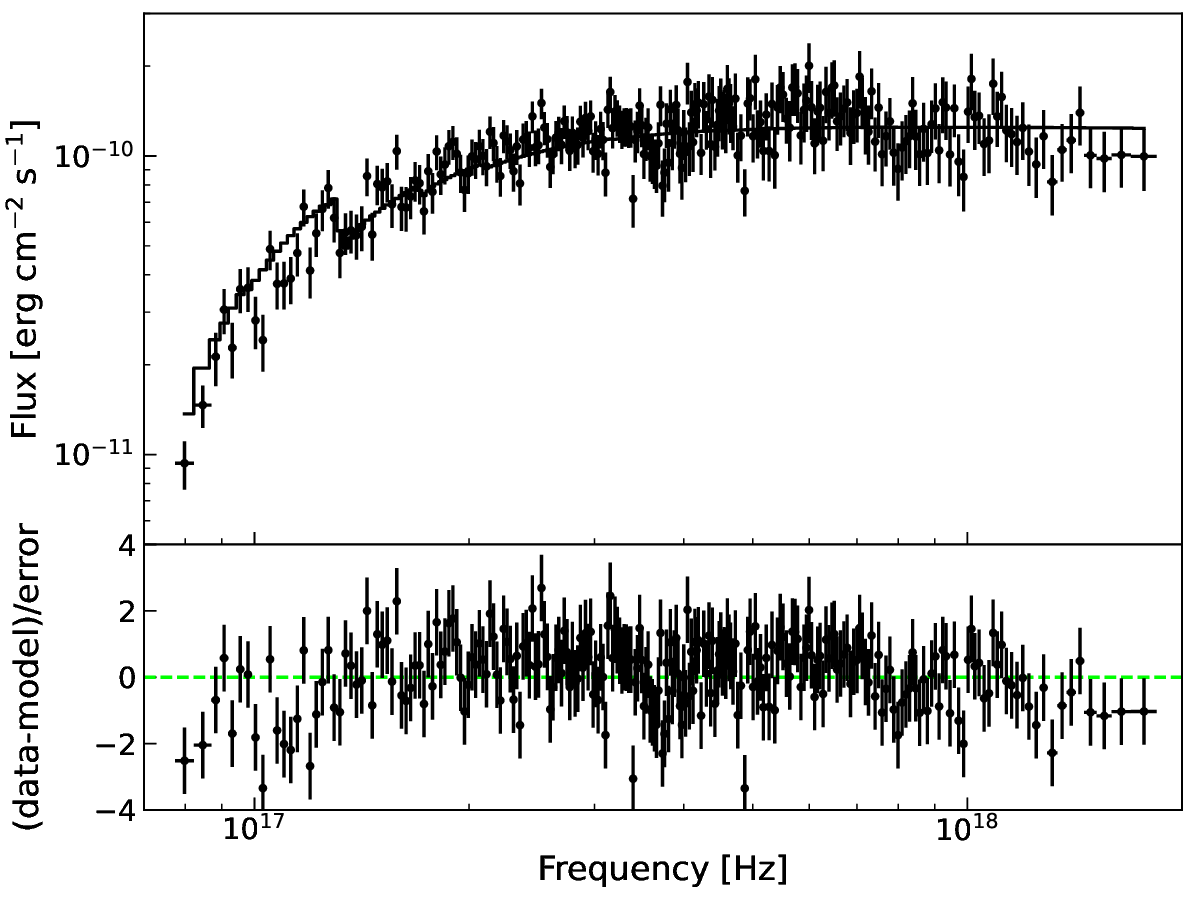}
    \includegraphics[angle=0,scale=0.4]{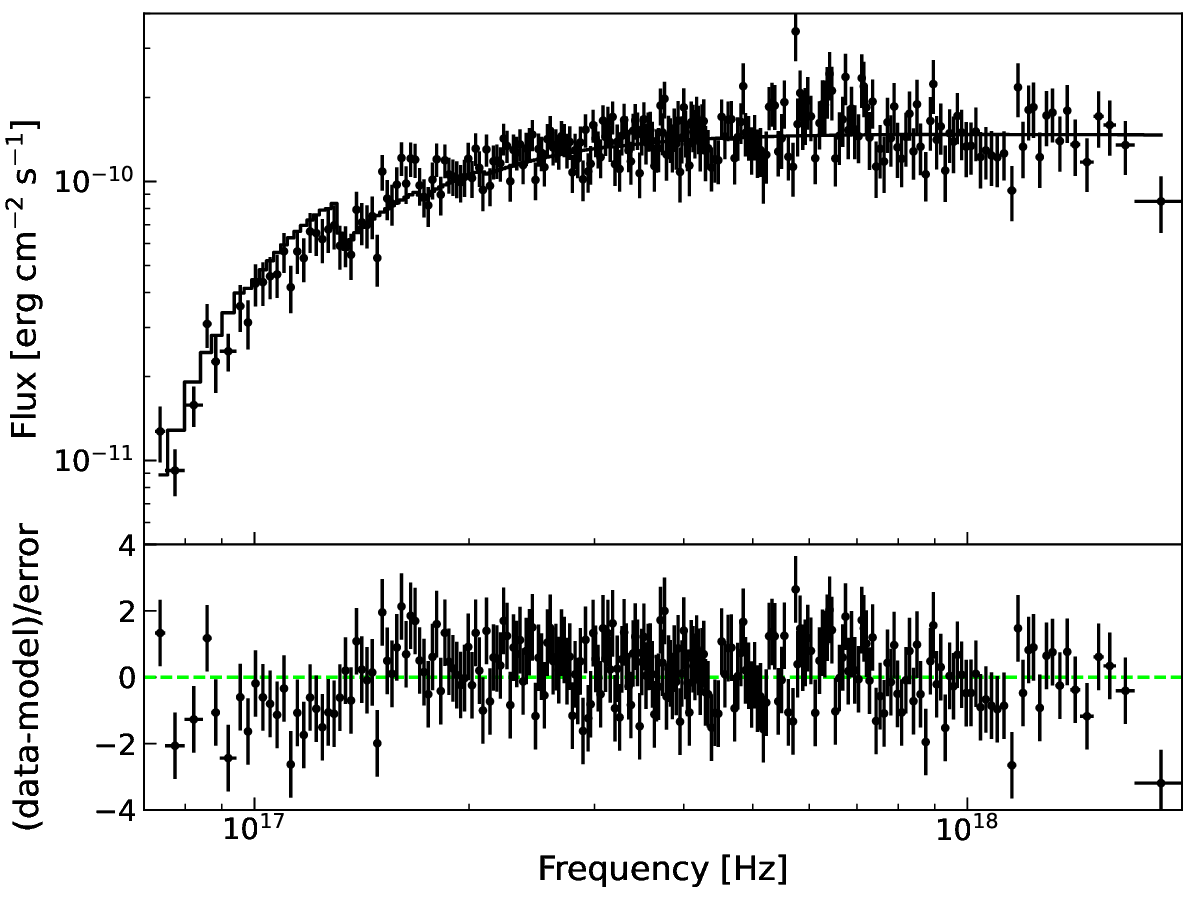}
    \includegraphics[angle=0,scale=0.4]{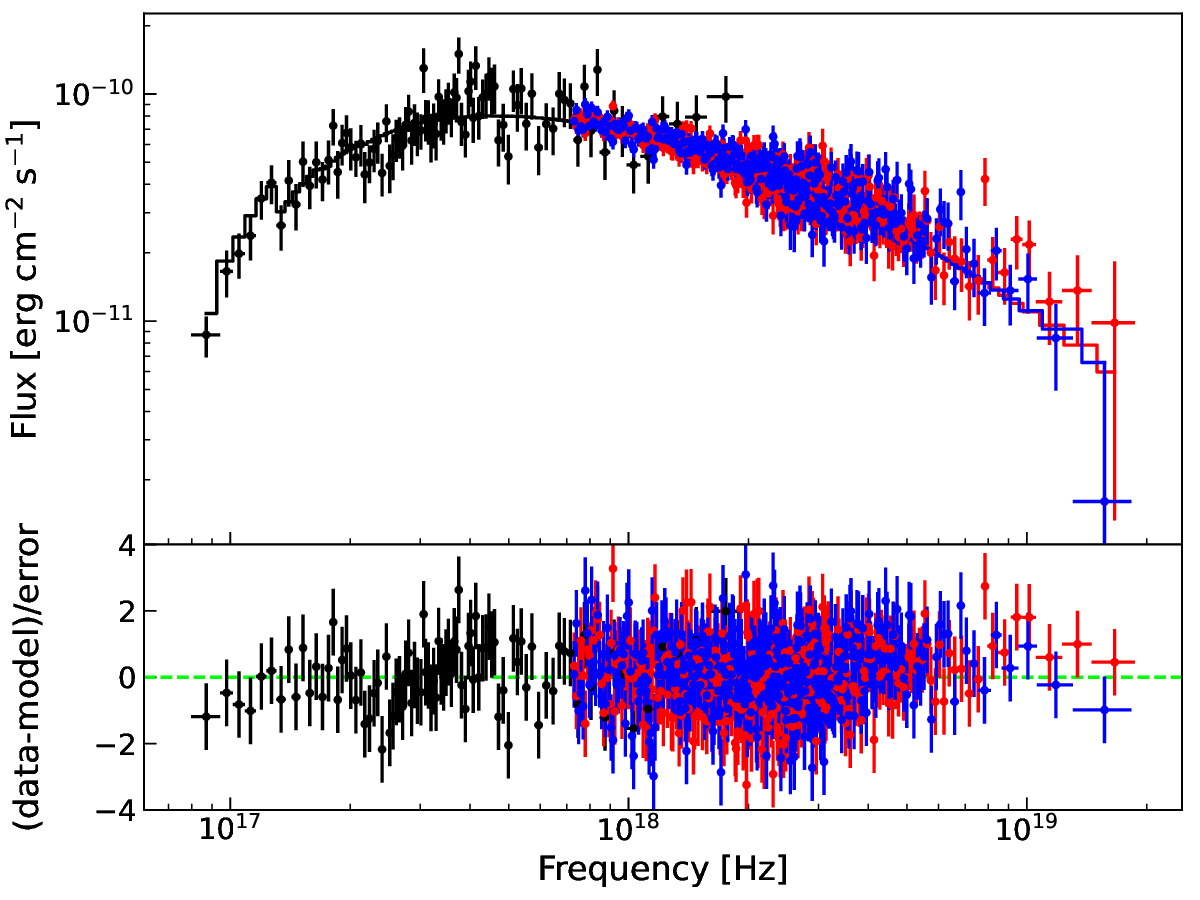}
    \includegraphics[angle=0,scale=0.4]{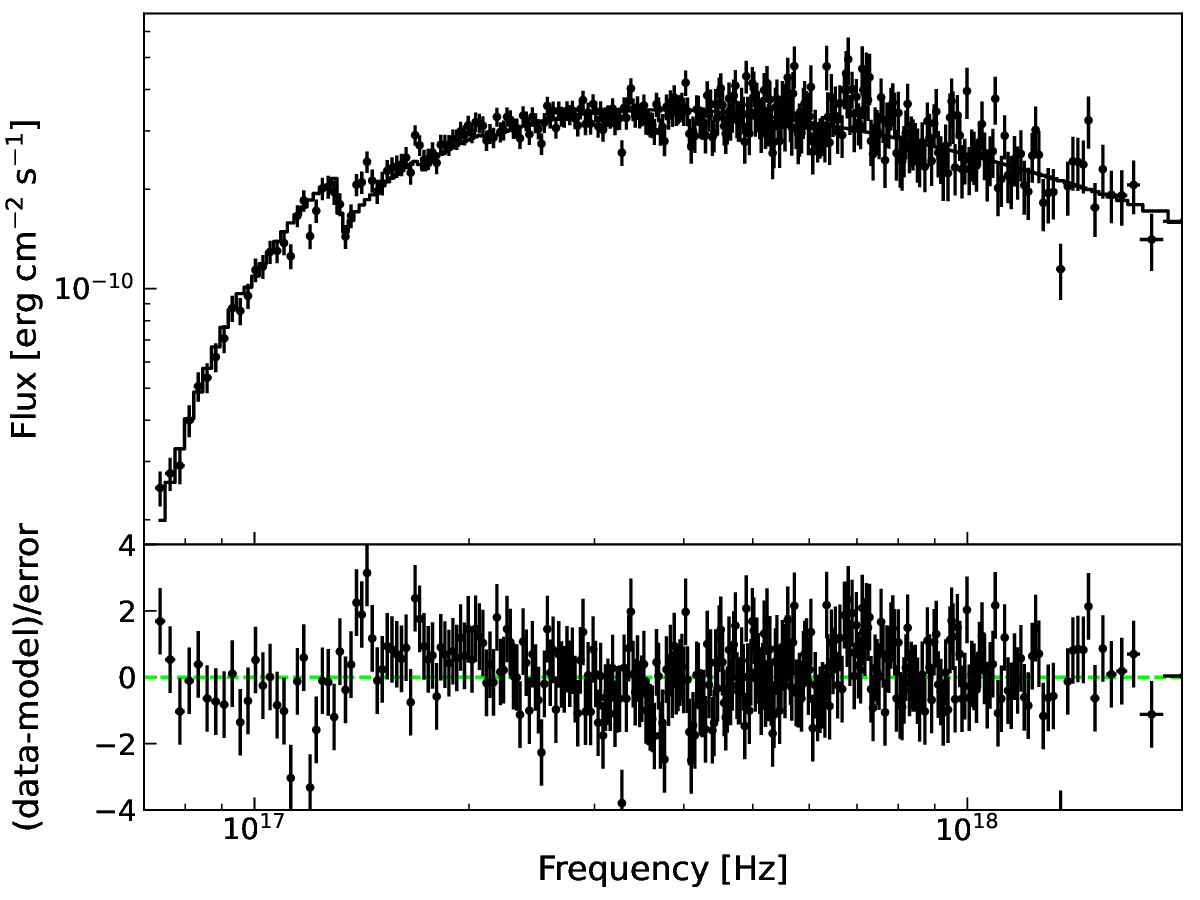}
    \includegraphics[angle=0,scale=0.4]{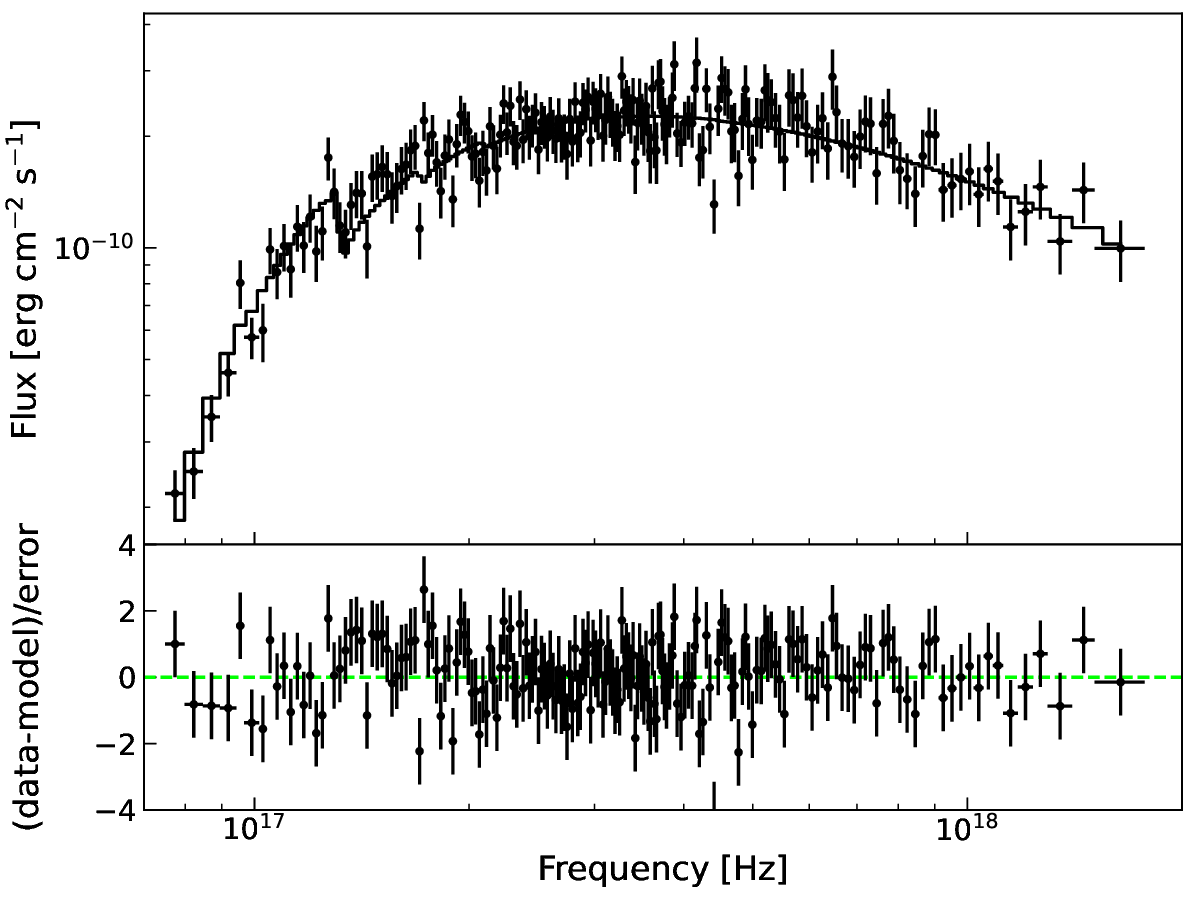}
    \includegraphics[angle=0,scale=0.4]{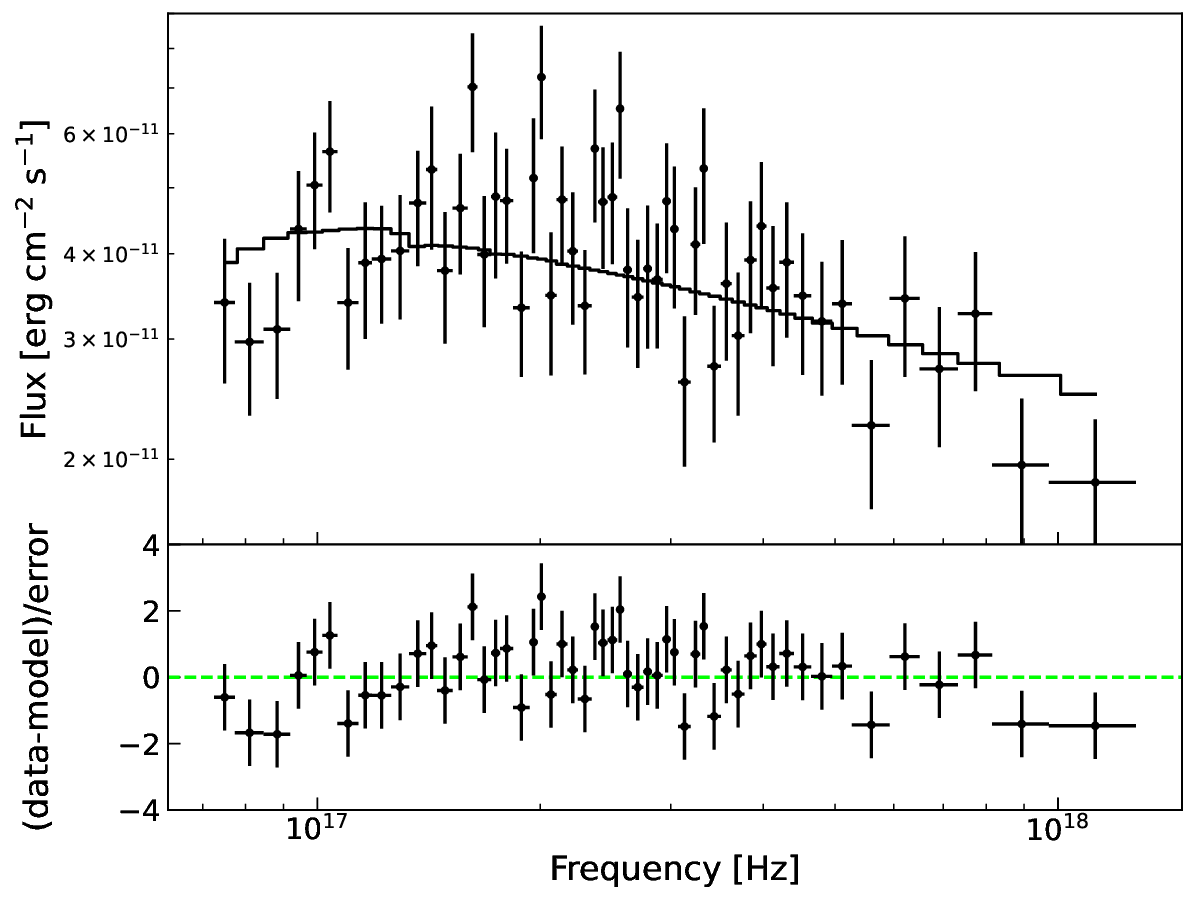}
    \caption{Spectral analysis results of X-rays. The black, red, and blue points indicate the data from Swift--XRT, NuSTAR--FPMA, and NuSTAR--FPMB, respectively. The six spectra relate to the observations performed on 2022 May 3 (top left), 2022 June 12 (top right), 2022 October 31 (mid left), 2023 August 14 (mid right), 2023 August 19 (bottom left), and 2023 November 1 (bottom right), respectively.}\label{XRT}
\end{figure}

\begin{figure}
    \centering
    \includegraphics[angle=0,scale=0.6]{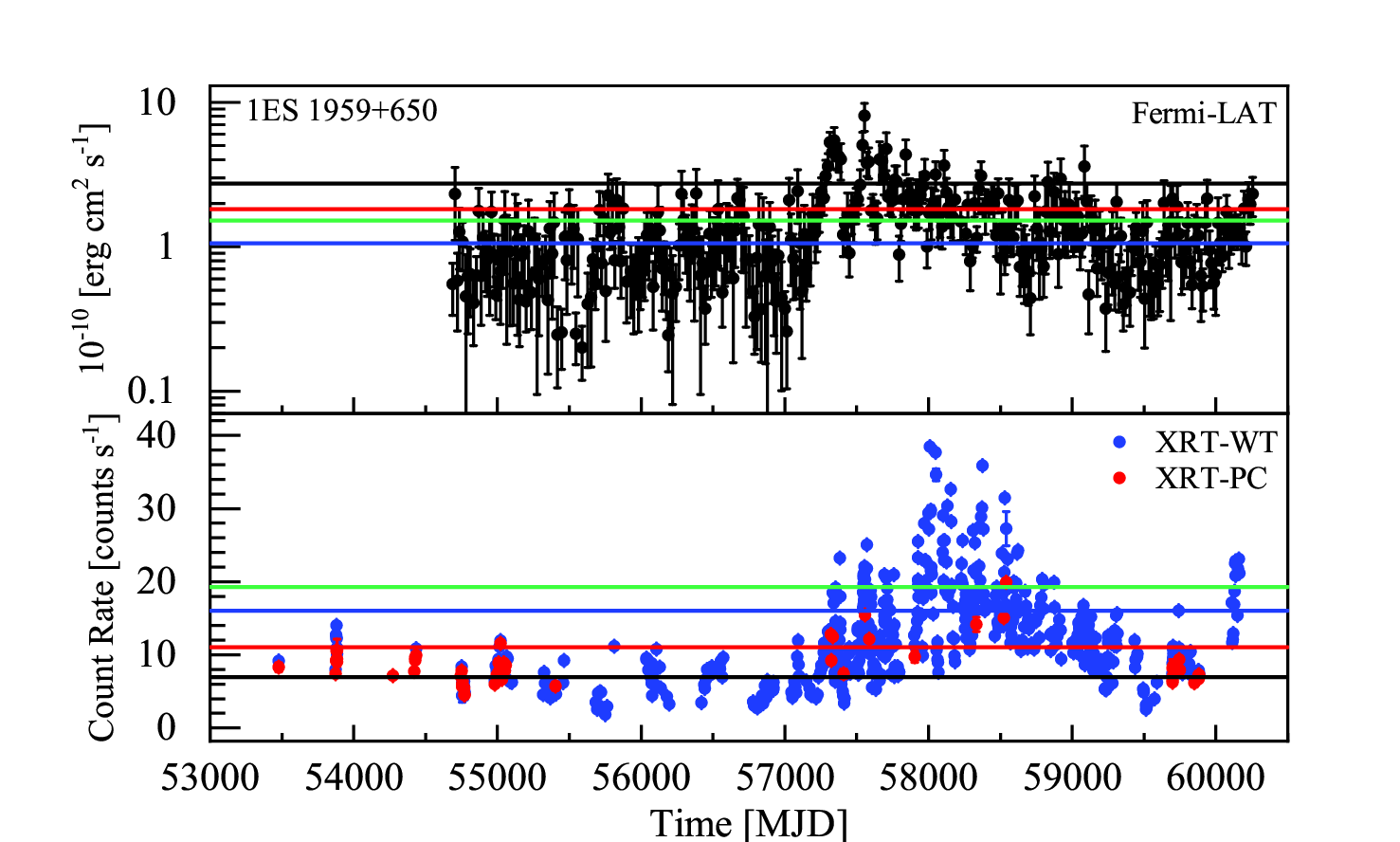}
    \includegraphics[angle=0,scale=0.6]{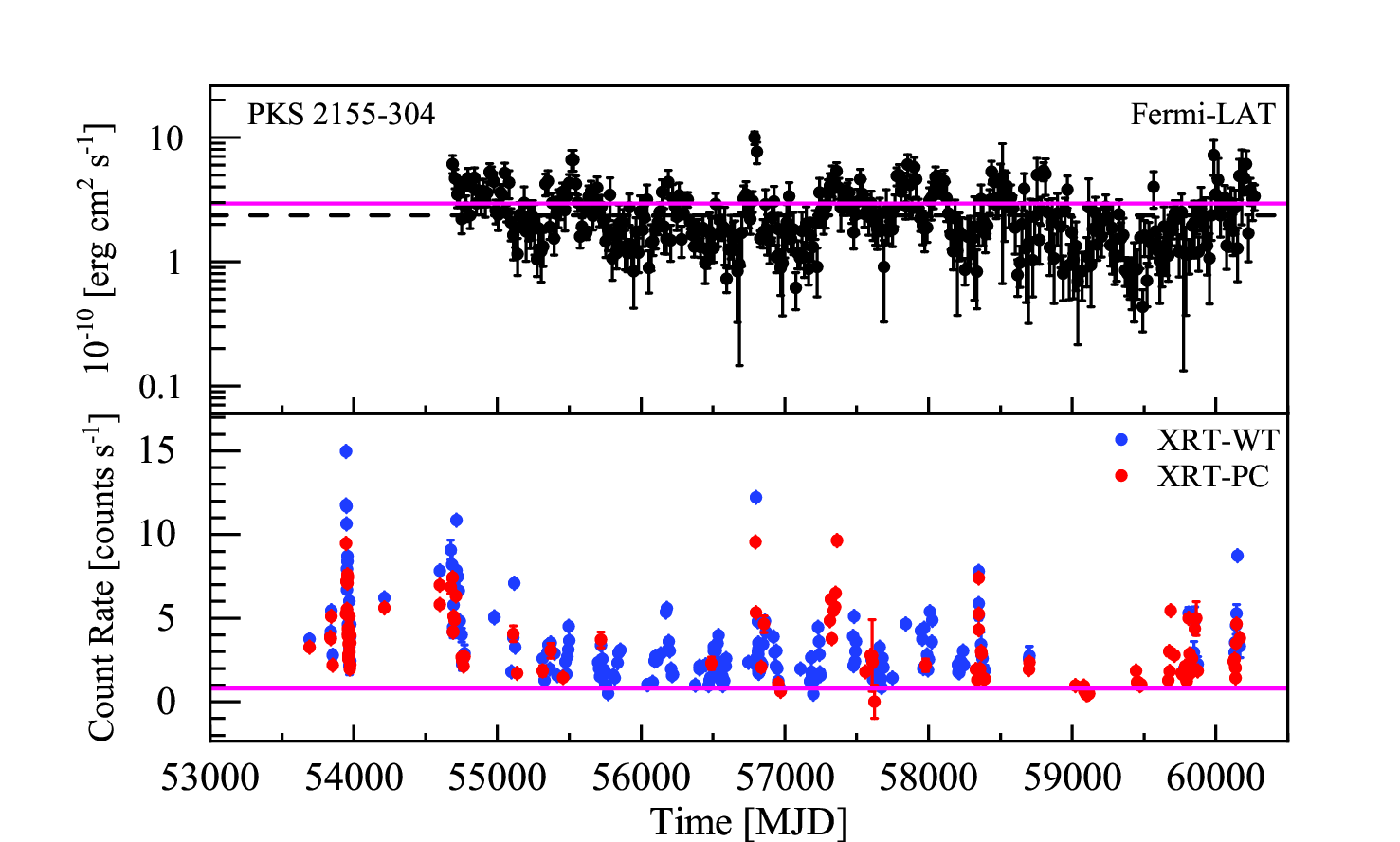}
    \caption{The long-term light curves derived by the Fermi--LAT and Swift--XRT observations for the two HBLs. The $\sim$15 yr $\gamma$-ray light curves are derived with an adaptive-binning method based on a criterion of TS>25 for each time bin, where the minimum time-bin step is 15 days. The XRT data are taken from a long-term Swift monitoring program of Fermi $\gamma$-ray sources \citep{2013ApJS..207...28S}. In the top panel, the red, blue, black, and green horizontal lines indicate the flux states in X-ray and $\gamma$-ray bands during the first, second, third, and fourth IXPE observations for 1ES 1959+650. In the bottom panel, the magenta horizontal lines indicate the flux states in X-ray and $\gamma$-ray bands during the IXPE observation for PKS 2155--304, while the black dashed horizontal line represents the $\sim$15 yr average flux in the Fermi--LAT band.}\label{LC}
\end{figure}

\begin{figure}
    \centering
    \includegraphics[angle=0,scale=0.35]{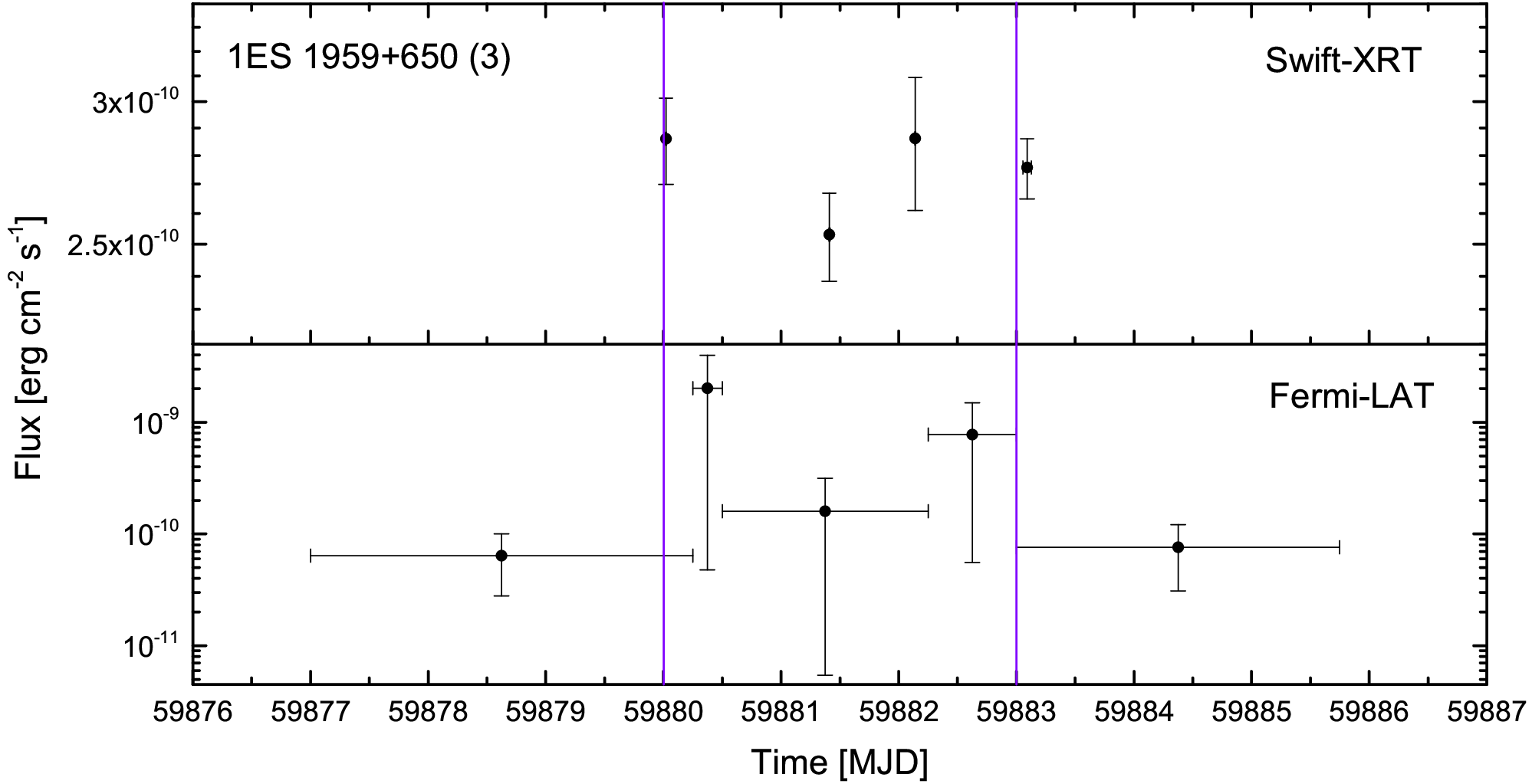}
    \includegraphics[angle=0,scale=0.35]{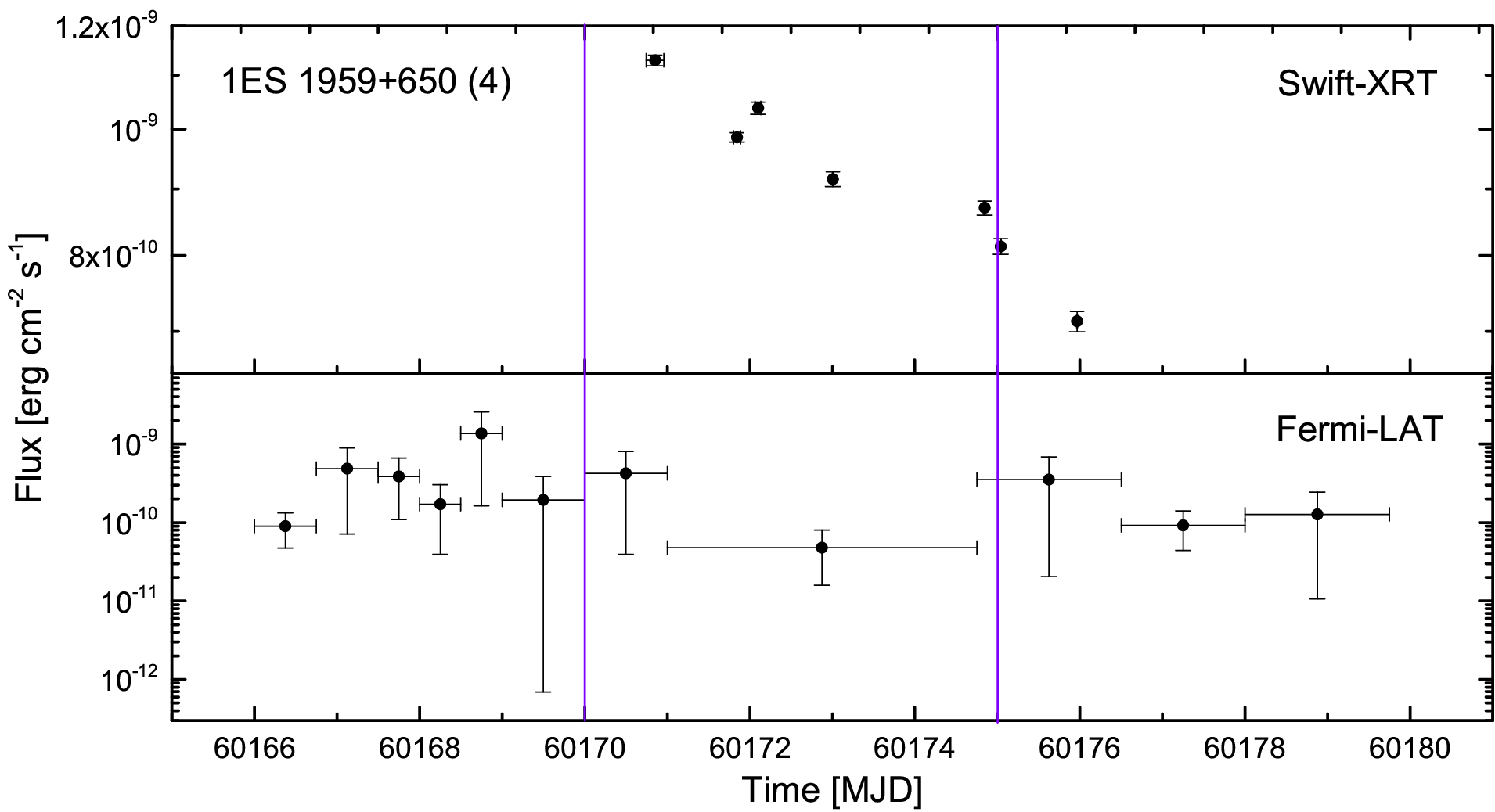}
    \includegraphics[angle=0,scale=0.35]{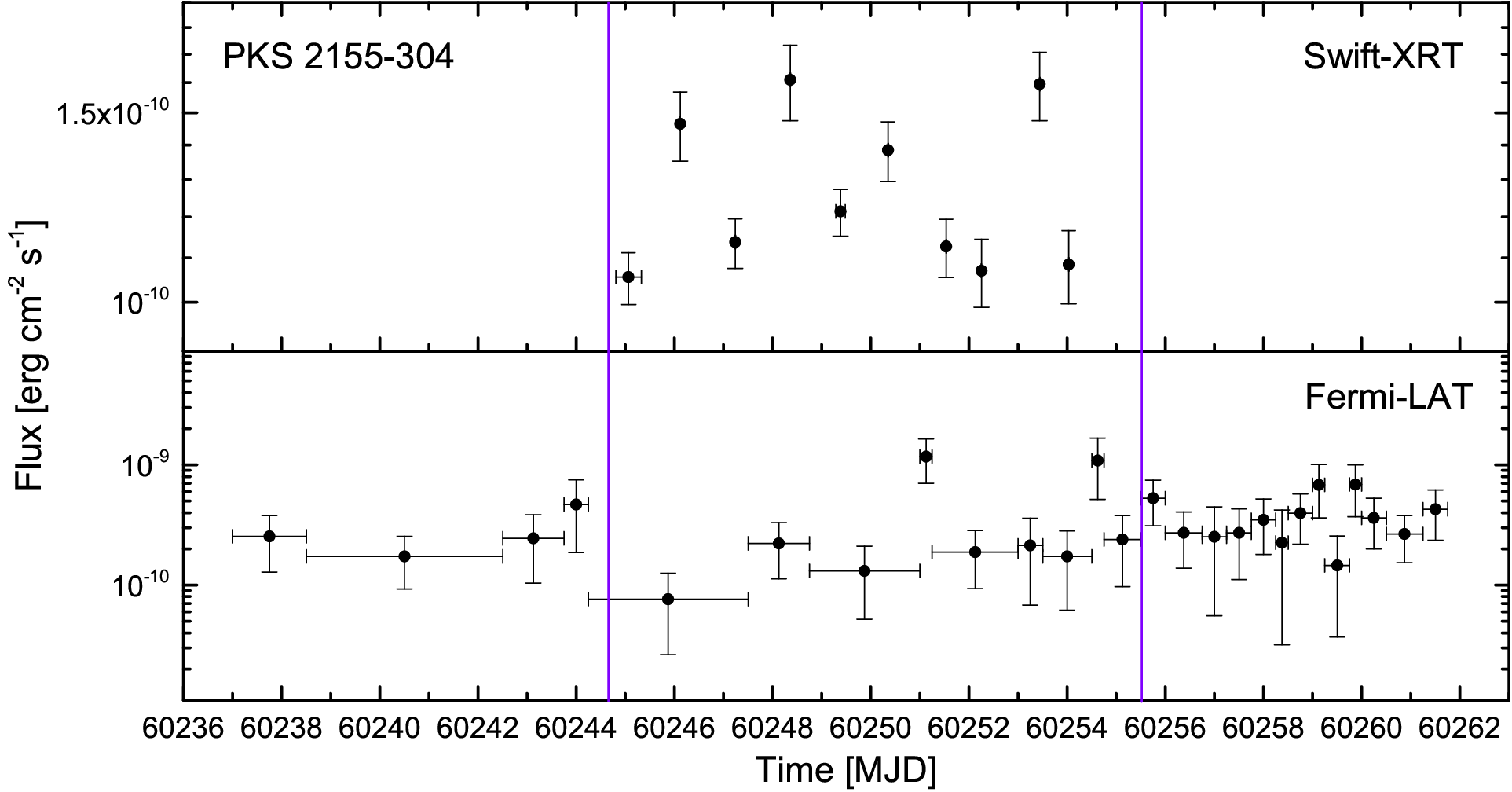}
    \caption{The light curves in X-ray and $\gamma$-ray bands during the IXPE observations for the two HBLs. The X-rays are obtained by analysing the XRT observational data. The $\gamma$-ray light curves are derived with an adaptive-binning method based on a criterion of TS>25 for each time bin, where the minimum time-bin step is 6 hr. The two purple vertical lines represent the IXPE observation time. The top and middle panels respectively correspond to the third and fourth IXPE observations of 1ES 1959+650. The bottom panel indicates the result during the IXPE observation of PKS 2155--304.}\label{LC-IXPE}
\end{figure}

\begin{figure}
    \centering
    \includegraphics[angle=0,scale=0.50]{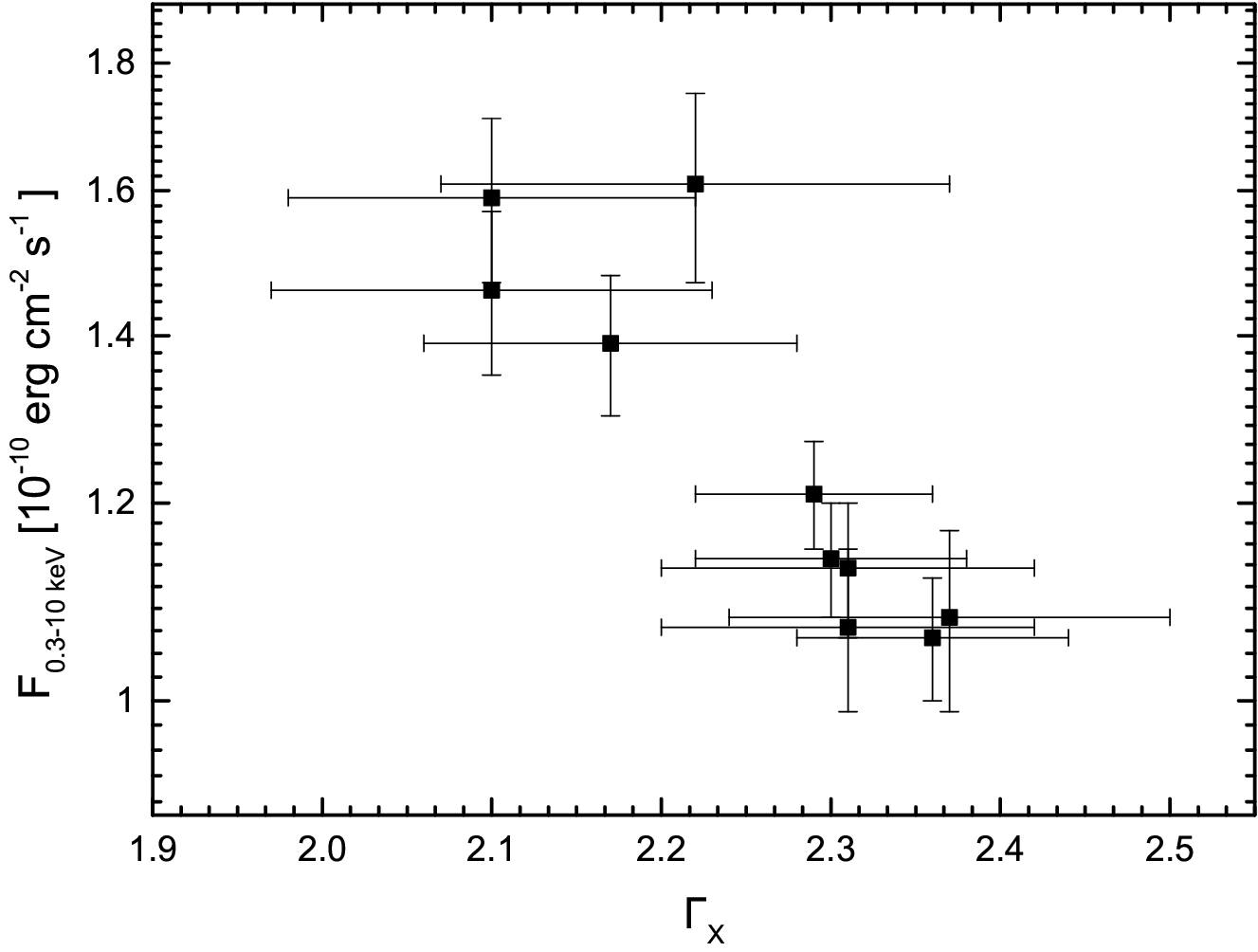}
    \caption{$\Gamma_{\rm X}$ against $F_{\rm 0.3-10~keV}$ for PKS 2155--304, where $\Gamma_{\rm X}$ and $F_{\rm 0.3-10~keV}$ are the photon spectral index and flux in the 0.3--10 keV band and derived with the XRT observation data during the IXPE observation. Using the bootstrap method to estimate the correlation coefficient ($r$), we obtain $r=-0.54$.}\label{F-G_2155}
\end{figure}

Polarimetry measurements are a valuable tool for investigating radiation mechanisms. For example, by combining polarimetry measurement in the optical band conducted by HST observations with spectra obtained from both optical and X-ray bands \citep{2006ApJ...651..735P,2020Galax...8...71P,2013ApJ...773..186C}, it is possible to determine the origin of X-ray emission from large-scale jet substructures, being due to synchrotron or inverse-Comptonized cosmic microwave background. Additionally, X-ray polarimetry can be utilized as a probe to investigate the X-ray emission mechanisms, including synchrotron, SSC or external-Compton \citep[e.g.,][]{2012ApJ...744...30K,2019ApJ...885...76P}. The disappearance of polarization in the high energy range observed by IXPE may provide insights into the transition region between synchrotron and SSC \citep[syc--SSC transition,][]{2019ApJ...885...76P}, as demonstrated by IXPE observations of BL Lacertae \citep{2023ApJ...948L..25P}. The first and third IXPE observations of 1ES 1959+650, as well as the IXPE observation of PKS 2155--304, showed no significant polarization above 5 keV (Figure \ref{IXPE_kev}), similar to BL Lacertae. However, when combining the XRT and NuSTAR observational data, there is no evidence of a harder spectrum at high-energy X-ray bands for 1ES 1959+650 during its third IXPE observation like that seen in BL Lacertae. Therefore, the absence of polarization above 5 keV should not be attributed to the syn--SSC transition. In contrast, for PKS 2155--304, higher polarization appears to be present in higher energy bins, together with a steep spectrum and an unusually low flux level during the IXPE observation, suggesting that the tail of synchrotron emission falls within the IXPE energy band. The synchrotron emission tail is necessarily dominated by a few zones and a high polarization is expected \citep{2019ApJ...885...76P}, thus indicating that vanishing polarization above 5 keV in PKS 2155--304 may indeed be due to the syn--SSC transition. Nevertheless, we cannot draw more accurate conclusions at this point due to insufficient hard X-ray data and other multiwavelength observations, especially regarding the first IXPE observation of 1ES1959+650. Additionally, no significant polarization was detected in the highest energy bin for both sources (Figure \ref{IXPE_kev}), except for the fourth IXPE observation of 1ES 1959+650 during a period of high-flux X-ray emission from the source. This could be attributed to the significant decrease in the effective detection area of IXPE in the high-energy range (in particular above 7 keV) within the 6--8 keV band (see Figure 5.1 in IXPE User's Guide: Observatory\footnote{\url{https://heasarc.gsfc.nasa.gov/docs/ixpe/analysis/IXPE_SOC_DOC_011-UserGuide-Observatory.pdf}}).

Four IXPE observations were performed for 1ES 1959+650 and the source exhibits the evident X-ray polarization variability accompanied by the variations of $\psi_{\rm X}$, spectrum, and flux, and energy bin. The estimated X-ray polarization degrees are higher than the archival optical and radio polarization degrees reported on the website\footnote{\url{http://www.bu.edu/blazars/VLBAproject.html}}. However, simultaneous polarimetry data in the optical and radio bands are not available. The derived values of $\psi_{\rm X}$ are $\psi_{\rm X}=129.9\degr\pm9.3\degr$, $\psi_{\rm X}=53.4\degr\pm4.9\degr$, and $\psi_{\rm X}=19.7\degr\pm1.6\degr$ for the first, third, and fourth IXPE observations, respectively. They are not all consistent with the jet axis projection on the plane of the sky, which is $127.9\degr\pm12.9\degr$ \citep{2022ApJS..260...12W}. A clear rotation of the X-ray polarization angle was presented. Recently, a smooth orphan optical polarization rotation was observed during IXPE observation in the HBL PG 1553+113 \citep{2023ApJ...953L..28M}, which was attributed to turbulence based on simulation results from the Turbulent Extreme MultiZone and Particle-in-Cell models \citep{2017Galax...5...63M,2021Galax...9...27M,2023ApJ...949...71Z}. The rotation of X-ray polarization angle in 1ES 1959+650 was accompanied by the variations in flux, spectra, and polarization degree. Based on the limited observation data, we speculate that the rotation of X-ray polarization angle in 1ES 1959+650 may be associated with an emission feature propagating along a helical magnetic field \citep[e.g.,][]{2008Natur.452..966M,2012ApJ...744...30K,2016MNRAS.457.2252B}.

However, due to the long time interval between the four IXPE observations of 1ES 1959+650 and the observed variations in flux, spectrum, and polarization during these observations, we believe that the X-rays during these observations originate from separate regions in the jet, and the acceleration mechanisms of particles may also differ. For the first IXPE observation of 1ES 1959+650, the estimated $\psi_{\rm X}$ in the 2--4 keV band was well align with the jet axis, as observed in Mrk 501 \citep{2022Natur.611..677L}, likely implying that the emission feature encounters a shock and particles are accelerated by it \citep[e.g.,][]{2023ApJ...948L..25P}. During the second IXPE observation, no polarization was detected across any energy bin, and the simultaneous Fermi--LAT observations indicated a low $\gamma$-ray emission state with a small TS$\sim$18. We speculate that these detected X-rays may originate from a large volume with disordered magnetic fields. During both the third and fourth IXPE observations, the source exhibited a log-parabola spectral form in the X-ray band, a characteristic feature commonly seen in synchrotron-dominated HBLs \citep[e.g.,][]{2004A&A...413..489M,2021MNRAS.507.5690G,2022MNRAS.514.3179M}. The curved log-parabola spectra of the synchrotron emission may be a fingerprint of stochastic acceleration \citep{2008A&A...489.1047M,2011ApJ...739...66T}. Additionally, derived values of $\psi_{\rm X}$ indicate that the EVPA orientation was nearly orthogonal to the jet axis. Therefore, a few zones with turbulent magnetic fields may dominate the X-ray emission and result in the high polarization levels and the random EVPA orientations \citep[e.g.,][]{2014ApJ...780...87M,2018ApJ...864..140P,2023ApJ...948L..25P} for the last two IXPE observations of 1ES 1959+650.

The IXPE observation of PKS 2155--304 was conducted during a period of historically low X-ray flux, however, significant polarization was observed, i.e., $\Pi_{\rm X}=21.9\%\pm1.9\%$ (MDP$_{99}\sim$6.0\%) and $\psi_{\rm X}=129.9\degr\pm2.5\degr$ in the 2--8 keV band and $\Pi_{\rm X}=28.6\%\pm2.7\%$ (MDP$_{99}\sim$8.1\%) in the 3--4 keV band. The simultaneous XRT spectra can be well explained by a steep ($\Gamma_{\rm X}$>2) power-law function, indicating that the synchrotron emission peak lies below the IXPE energy range while its high-energy tail extends into this range. We checked the archived optical polarization data of PKS 2155--304 from the Steward Observatory spectropolarimetric monitoring project\footnote{\url{http://james.as.arizona.edu/~psmith/Fermi/}} \citep{2009arXiv0912.3621S} and found that the highest detected polarization in the optical band was $\Pi_{\rm O}=19.1\%\pm0.1\%$ with $\psi_{\rm O}=59.1\degr\pm0.1\degr$ on 2017 May 22. No radio polarization data or information about the direction of the jet axis for PKS 2155--304 was found in the literature. During the IXPE observation, the variations of X-ray flux were observed, which were accompanied by the spectral variations, displaying a tendency of \emph{harder when brighter} behavior. However, no clear flux variation was found in the simultaneously observed Fermi--LAT data. The \emph{harder when brighter} behavior is commonly observed among HBLs in the X-ray band and has also been documented for Mrk 421 \citep{2022ApJ...938L...7D}. This behavior is typically attributed to the shock-induced injection of high-energy electrons \citep{1998A&A...333..452K,2022ApJ...938L...7D}. Therefore, similar to Mrk 421 and Mrk 501 \citep{2022ApJ...938L...7D,2022Natur.611..677L}, the X-rays of PKS 2155--304 are produced by synchrotron radiation in a region with ordered magnetic fields and the particles are accelerated by a shock.

\acknowledgments

We thank the anonymous referee for valuable suggestions. This work is supported by the National Key R\&D Program of China (grant 2023YFE0117200) and the National Natural Science Foundation of China (grants 12022305, 11973050, 12373042, 12133003, U1938201, 12203022).

\clearpage

\bibliography{reference}

\clearpage

\begin{sidewaystable}
    \scriptsize
    \setlength{\tabcolsep}{1.5pt}
    \centering
    \caption{Analysis Results of IXPE Data for both 1ES 1959+650 and PKS 2155--304}
    \begin{center}
        \begin{tabular}{ccccccccccccccccccccc}
            \hline
            \hline
            Method\footnote{The methods used to estimate the polarization. ``I'' indicates model-independent polarimetry using the software $ixpeobssim$, while ``II'' indicates the spectropolarimetric fitting using Xspec. Please note that spectropolarimetric fitting is only performed when the derived value of $\Pi_{\rm X}$ with the software $ixpeobssim$ exceeds MDP$_{99}$.} & Object\footnote{The name of sources. ``1959'' and ``2155'' denote 1ES 1959+650 and PKS 2155--304, respectively.} & OBSID\footnote{The unique identification number specifying the IXPE observation. ``A'' for 01006201 with exposure of 53519 s, ``B'' for 01006001 with exposure of 200432 s, ``C'' for 02004801 with exposure of 184844 s, ``D'' for 02250801 with exposure of 312447 s, and ``E'' for 02005601 with exposure of 476108 s.} & Date\footnote{The start time of the IXPE observation.} & $\Pi_{\rm X}$ & $\psi_{\rm X}$ & MDP$_{99}$ & $\Pi_{\rm X}$ & MDP$_{99}$ &$\Pi_{\rm X}$ & MDP$_{99}$ & $\Pi_{\rm X}$ & MDP$_{99}$ & $\Pi_{\rm X}$ & MDP$_{99}$ & $\Pi_{\rm X}$ & MDP$_{99}$ & $\Pi_{\rm X}$ & MDP$_{99}$ \\
            & & & & (\%) & ($\degr$) & (\%) & (\%) & (\%) & (\%) & (\%)& (\%) & (\%) & (\%) & (\%) & (\%) & (\%) & (\%) & (\%) \\
            & & & & 2--8 keV & 2--8 keV & 2--8 keV & 2--3 keV & 2--3 keV & 3--4 keV & 3--4 keV & 4--5 keV & 4--5 keV & 5--6 keV & 5--6 keV & 6--7 keV & 6--7 keV & 7--8 keV & 7--8 keV \\
            \hline
            I & 1959 & A & 2022--05--03 & $6.0\pm2.4$ & \nodata & 7.5 & $9.4\pm3.1$ & 9.3 & $9.1\pm3.6$ & 10.9 & $4.7\pm5.8$ & 17.5 & $9.1\pm9.0$ & 27.4 & $24.6\pm14.9$ & 45.2 & $62.8\pm29.1$ & 87.4 \\
            & & B  & 2022--06--09 & $2.1\pm1.2$ & \nodata & 3.8 & $2.1\pm1.5$ & 4.7 & $2.5\pm1.7$ & 5.3 & $3.9\pm2.8$ & 8.5 & $3.2\pm4.3$ & 13.1 & $6.9\pm6.6$ & 19.9 & $2.5\pm12.5$ & 38.0 \\
            & & C & 2022--10--28 & $9.3\pm1.6$ & $53.4\pm4.9$ & 4.9 & $8.1\pm2.0$ & 6.1 & $12.7\pm2.3$ & 7.0 & $12.4\pm3.8$ & 11.5 & $2.2\pm6.0$ & 18.3 & $12.4\pm9.4$ & 28.6 & $30.0\pm21.5$ & 65.1 \\
            & & D & 2023--08--14 & $12.4\pm0.7$ & $19.7\pm1.6$ & 2.2 & $11.5\pm0.9$ & 2.7 & $11.1\pm1.1$ & 3.3 & $15.9\pm1.8$ & 5.5 & $10.4\pm2.9$ & 8.8 & $16.2\pm4.7$ & 14.2 & $37.0\pm9.3$ & 28.0 \\
            & 2155 & E  & 2023--10--27 & $21.9\pm1.9$ & $129.9\pm2.5$ & 6.0 & $21.0\pm2.1$ & 6.5 & $28.6\pm2.7$ & 8.1 & $20.4\pm4.6$ & 14.1 & $20.5\pm7.9$ & 24.1 & $10.0\pm13.6$ & 41.3 & $11.2\pm31.5$ & 95.6 \\
            \hline
            II & 1959& A  & 2022--05--03 & \nodata & \nodata & \nodata & $8.0\pm3.2$ & \nodata & \nodata & \nodata & \nodata & \nodata & \nodata & \nodata & \nodata & \nodata & \nodata & \nodata \\
            & & B  & 2022--06--09 & \nodata & \nodata & \nodata & \nodata & \nodata & \nodata & \nodata & \nodata & \nodata & \nodata & \nodata & \nodata & \nodata & \nodata & \nodata \\
            & & C & 2022--10--28 & $9.8\pm1.4$ & $51.3\pm4.1$ & \nodata & $7.8\pm2.1$ & \nodata & $13.6\pm2.4$ & \nodata & $13.7\pm3.8$ & \nodata & \nodata & \nodata & \nodata & \nodata & \nodata & \nodata \\
            & & D & 2023--08--14 & $12.5\pm0.1$ & $20.1\pm1.6$ & \nodata & $12.5\pm0.9$ & \nodata & $11.4\pm1.1$ & \nodata & $15.6\pm1.8$ & \nodata & $9.8\pm3.0$ & \nodata & $16.5\pm4.8$ & \nodata & $30.1\pm9.2$ & \nodata \\
            & 2155 & E  & 2023--10--27 & $24.1\pm1.6$ & $129.4\pm1.9$ & \nodata & $22.5\pm2.3$ & \nodata & $29.7\pm2.9$ & \nodata & $21.0\pm4.7$ & \nodata & \nodata & \nodata & \nodata & \nodata & \nodata & \nodata \\
            \hline
        \end{tabular}
        \label{table1}
    \end{center}
\end{sidewaystable}

\begin{deluxetable}{cccccccc}
    \tabletypesize{\footnotesize}
    \tablecolumns{8}
    \tablecaption{Swift--XRT and NuSTAR Observations During Four IXPE Observations of 1ES 1959+650}
    \tablehead{\colhead{OBSID} & \colhead{Date} & \colhead{Exp Time} & \colhead{Mode\tablenotemark{\footnotesize{a}}} & \colhead{$\Gamma_{\rm X}$} & \colhead{$\beta$} & \colhead{Flux\tablenotemark{\footnotesize{b}}} & \colhead{$\chi^{\rm 2}$/dof} \\
    \colhead{} & \colhead{} & \colhead{(s)} & \colhead{} & \colhead{} & \colhead{} & \colhead{(10$^{-10}$ erg cm$^{-2}$ s$^{-1}$)} & \colhead{}}
    \startdata
    00096560006 & 2022--05--03 & 909 & WT & $2.02\pm0.03$ & 0 & $4.50\pm0.08$ & 289/251 \\
    00096560007 & 2022--05--04 & 884 & WT & $1.99\pm0.03$ & 0 & $4.74^{+0.08}_{-0.09}$ & 321/251 \\
    00096560012 & 2022--06--12 & 934 & WT & $2.02\pm0.03$ & 0 & $5.29\pm0.10$ & 253/245 \\
    00096560022 & 2022--10--28 & 920 & PC & $2.04\pm0.09$ & 0 & $2.86^{+0.15}_{-0.16}$ & 49/46 \\
    00096560023 & 2022--10--29 & 910 & PC & $1.94\pm0.09$ & 0 & $2.53^{+0.14}_{-0.15}$ & 49/43 \\
    00096560024 & 2022--10--30 & 725 & PC & $1.93\pm0.16$ & 0 & $2.86^{+0.23}_{-0.25}$ & 29/20 \\
    00089446001 & 2022--10--31 & 1729 & PC & $1.98\pm0.05$ & 0 & $2.76\pm0.10$ & 97/90 \\
    00097164002 & 2023--08--14 & 1715 & WT & $2.08\pm0.02$ & $0.40\pm0.04$ & $11.30\pm0.10$ & 377/332 \\
    00013906083 & 2023--08--15 & 1693 & WT & $2.12\pm0.02$ & $0.37\pm0.04$ & $9.86\pm0.08$ & 425/347 \\
    00013906084 & 2023--08--16 & 1037 & WT & $1.98\pm0.02$ & $0.52\pm0.05$ & $10.40\pm0.11$ & 368/307 \\
    00097164003 & 2023--08--17 & 773 & WT & $1.96\pm0.03$ & $0.60\pm0.07$ & $9.16\pm0.12$ & 294/256 \\
    00097164004 & 2023--08--18 & 1102 & WT & $2.17\pm0.03$ & $0.38\pm0.06$ & $8.70\pm0.11$ & 309/262 \\
    00097164005 & 2023--08--19 & 795 & WT & $2.16\pm0.03$ & $0.40\pm0.07$ & $8.13\pm0.11$ & 267/240 \\
    00013906085 & 2023--08--19 & 885 & WT & $2.09\pm0.04$ & $0.50\pm0.09$ & $7.12\pm0.13$ & 188/192 \\
    \hline
    60801022002 & 2022--10--31 & 17815 & FPMA & $2.30^{+0.14}_{-0.15}$ & $0.18^{+0.09}_{-0.08}$ & $1.16\pm0.01$ & 331/323 \\
    & & 17649 & FPMB & $2.1\pm0.15$ & $0.27^{+0.09}_{-0.08}$ & $1.16\pm0.01$ & 358/312 \\
    & & 17732 & FPMA+FPMB & $2.21^{+0.10}_{-0.11}$ & $0.23\pm0.06$ & $1.16\pm0.01$ & 693/638 \\
    \enddata
    \tablenotetext{a}{WT and PC represent the readout mode of Swift--XRT, while FPMA and FPMB represent the two multilayer-coated telesocpes of NuSTAR.}
    \tablenotetext{b}{Flux is calculated in the 0.3--10 keV band for Swift--XRT data and in the 3--79 keV band for NuSTAR data, respectively.}
    \label{table2}
\end{deluxetable}

\begin{deluxetable}{ccccccc}
    \tabletypesize{\small}
    \tablecolumns{7}
    \tablecaption{Swift--XRT Observations During the IXPE Observation of PKS 2155--304}
    \tablehead{\colhead{OBSID} & \colhead{Date} & \colhead{Exp Time} & \colhead{Mode} & \colhead{$\Gamma_{\rm X}$} & \colhead{Flux} & \colhead{$\chi^{\rm 2}$/dof} \\
    \colhead{} & \colhead{} & \colhead{(s)} & \colhead{} & \colhead{} & \colhead{(10$^{-10}$ erg cm$^{-2}$ s$^{-1}$)} & \colhead{}}
    \startdata
    00097172006 & 2023--10--27 & 1576 & PC & $2.36\pm0.08$ & $1.06\pm0.06$ & 50/42 \\
    00097172007 & 2023--10--29 & 892 & PC & $2.10\pm0.13$ & $1.46^{+0.10}_{-0.11}$ & 20/25 \\
    00097172008 & 2023--10--30 & 1211 & PC & $2.30\pm0.08$ & $1.14\pm0.06$ & 41/46 \\
    00097172009 & 2023--10--31 & 782 & PC & $2.22\pm0.15$ & $1.61^{+0.13}_{-0.14}$ & 11/21 \\
    00097172010 & 2023--11--01 & 1754 & PC & $2.29\pm0.07$ & $1.21\pm0.06$ & 55/52 \\
    00097172011 & 2023--11--02 & 907 & PC & $2.17\pm0.11$ & $1.39\pm0.09$ & 39/34 \\
    00097172012 & 2023--11--03 & 995 & PC & $2.31\pm0.11$ & $1.13\pm0.07$ & 35/34 \\
    00097172013 & 2023--11--04 & 899 & PC & $2.31\pm0.11$ & $1.07^{+0.07}_{-0.08}$ & 23/26 \\
    00097172014 & 2023--11--05 & 912 & PC & $2.10\pm0.12$ & $1.59^{+0.11}_{-0.12}$ & 24/26 \\
    00097172015 & 2023--11--06 & 853 & PC & $2.37\pm0.13$ & $1.08^{+0.08}_{-0.09}$ & 33/22 \\
    \enddata
\end{deluxetable}\label{table3}

\begin{deluxetable}{cccccc}
    \tabletypesize{\small}
    \tablecolumns{6}
    \tablecaption{Analysis Results of the Fermi--LAT Data for Two HBLs }
    \tablehead{\colhead{Object} & \colhead{Date} & \colhead{$\Gamma_{\gamma}$} & \colhead{$\beta$} & \colhead{Flux} & \colhead{TS} \\
    \colhead{} & \colhead{} & \colhead{} & \colhead{} & \colhead{(10$^{-11 }$ erg cm$^{-2}$ s$^{-1}$)} & \colhead{}}
    \startdata
    1ES 1959+650 &2008.08.04--2023.11.15 & $1.77\pm0.01$ & $0.015\pm0.003$ & $13.09\pm0.31$ & 31217.1 \\
    &2022.05.03--2022.05.04& $1.72\pm0.26$ & $0.015$\tablenotemark{\scriptsize{$*$}} & $18.24\pm12.95$ & 30.0 \\
    &2022.06.09--2022.06.12& $1.59\pm0.42$ & $0.015$\tablenotemark{\scriptsize{$*$}} & $10.59\pm10.13$ & 18.3 \\
    &2022.10.28--2022.10.31& $1.81\pm0.18$ & $0.015$\tablenotemark{\scriptsize{$*$}} & $27.43\pm12.75$ & 65.0 \\
    &2023.08.14--2023.08.19 & $1.79\pm0.20$ & $0.015$\tablenotemark{\scriptsize{$*$}} & $15.21\pm7.88$ & 49.8 \\
    \hline
    PKS 2155--304 &2008.08.04--2023.11.15 & $1.79\pm0.01$ & $0.027\pm0.003$ & $23.71\pm0.47$ & 78350.6 \\  
    &2023.10.27--2023.11.07 & $1.66\pm0.13$ & $0.027$\tablenotemark{\scriptsize{$*$}} & $29.48\pm10.00$ & 165.6 \\
    \enddata
    \tablenotetext{*}{It is fixed as the value of the $\sim$15 yr time-integrated spectrum.}
    \label{table4}
\end{deluxetable}

\clearpage

\end{document}